\documentclass[10pt,aps,prl,twocolumn,showpacs,superscriptaddress,floatfix,preprintnumbers,amsmath,amssymb]{revtex4-1}  

\usepackage{graphicx}  
\usepackage{dcolumn}   
\usepackage{bm}        
\usepackage{amssymb}   
\usepackage[pagewise]{lineno}
\usepackage[usenames,dvipsnames]{color}

\newcommand\etal{{\it {et al., }}}

\begin{document}

\preprint{APS/123-QED}

\title{Search for dark matter annihilations in the Sun with the 79-string IceCube detector}
 
\affiliation{III. Physikalisches Institut, RWTH Aachen University, D-52056 Aachen, Germany}
\affiliation{School of Chemistry \& Physics, University of Adelaide, Adelaide SA, 5005 Australia}
\affiliation{Dept.~of Physics and Astronomy, University of Alaska Anchorage, 3211 Providence Dr., Anchorage, AK 99508, USA}
\affiliation{CTSPS, Clark-Atlanta University, Atlanta, GA 30314, USA}
\affiliation{School of Physics and Center for Relativistic Astrophysics, Georgia Institute of Technology, Atlanta, GA 30332, USA}
\affiliation{Dept.~of Physics, Southern University, Baton Rouge, LA 70813, USA}
\affiliation{Dept.~of Physics, University of California, Berkeley, CA 94720, USA}
\affiliation{Lawrence Berkeley National Laboratory, Berkeley, CA 94720, USA}
\affiliation{Institut f\"ur Physik, Humboldt-Universit\"at zu Berlin, D-12489 Berlin, Germany}
\affiliation{Fakult\"at f\"ur Physik \& Astronomie, Ruhr-Universit\"at Bochum, D-44780 Bochum, Germany}
\affiliation{Physikalisches Institut, Universit\"at Bonn, Nussallee 12, D-53115 Bonn, Germany}
\affiliation{Universit\'e Libre de Bruxelles, Science Faculty CP230, B-1050 Brussels, Belgium}
\affiliation{Vrije Universiteit Brussel, Dienst ELEM, B-1050 Brussels, Belgium}
\affiliation{Dept.~of Physics, Chiba University, Chiba 263-8522, Japan}
\affiliation{Dept.~of Physics and Astronomy, University of Canterbury, Private Bag 4800, Christchurch, New Zealand}
\affiliation{Dept.~of Physics, University of Maryland, College Park, MD 20742, USA}
\affiliation{Dept.~of Physics and Center for Cosmology and Astro-Particle Physics, Ohio State University, Columbus, OH 43210, USA}
\affiliation{Dept.~of Astronomy, Ohio State University, Columbus, OH 43210, USA}
\affiliation{Dept.~of Physics, TU Dortmund University, D-44221 Dortmund, Germany}
\affiliation{Dept.~of Physics, University of Alberta, Edmonton, Alberta, Canada T6G 2G7}
\affiliation{D\'epartement de physique nucl\'eaire et corpusculaire, Universit\'e de Gen\`eve, CH-1211 Gen\`eve, Switzerland}
\affiliation{Dept.~of Physics and Astronomy, University of Gent, B-9000 Gent, Belgium}
\affiliation{Dept.~of Physics and Astronomy, University of California, Irvine, CA 92697, USA}
\affiliation{Laboratory for High Energy Physics, \'Ecole Polytechnique F\'ed\'erale, CH-1015 Lausanne, Switzerland}
\affiliation{Dept.~of Physics and Astronomy, University of Kansas, Lawrence, KS 66045, USA}
\affiliation{Dept.~of Astronomy, University of Wisconsin, Madison, WI 53706, USA}
\affiliation{Dept.~of Physics and Wisconsin IceCube Particle Astrophysics Center, University of Wisconsin, Madison, WI 53706, USA}
\affiliation{Institute of Physics, University of Mainz, Staudinger Weg 7, D-55099 Mainz, Germany}
\affiliation{Universit\'e de Mons, 7000 Mons, Belgium}
\affiliation{T.U. Munich, D-85748 Garching, Germany}
\affiliation{Bartol Research Institute and Department of Physics and Astronomy, University of Delaware, Newark, DE 19716, USA}
\affiliation{Dept.~of Physics, University of Oxford, 1 Keble Road, Oxford OX1 3NP, UK}
\affiliation{Dept.~of Physics, University of Wisconsin, River Falls, WI 54022, USA}
\affiliation{Oskar Klein Centre and Dept.~of Physics, Stockholm University, SE-10691 Stockholm, Sweden}
\affiliation{Department of Physics and Astronomy, Stony Brook University, Stony Brook, NY 11794-3800, USA}
\affiliation{Dept.~of Physics and Astronomy, University of Alabama, Tuscaloosa, AL 35487, USA}
\affiliation{Dept.~of Astronomy and Astrophysics, Pennsylvania State University, University Park, PA 16802, USA}
\affiliation{Dept.~of Physics, Pennsylvania State University, University Park, PA 16802, USA}
\affiliation{Dept.~of Physics and Astronomy, Uppsala University, Box 516, S-75120 Uppsala, Sweden}
\affiliation{Dept.~of Physics, University of Wuppertal, D-42119 Wuppertal, Germany}
\affiliation{DESY, D-15735 Zeuthen, Germany}

\author{M.~G.~Aartsen}
\affiliation{School of Chemistry \& Physics, University of Adelaide, Adelaide SA, 5005 Australia}
\author{R.~Abbasi}
\affiliation{Dept.~of Physics and Wisconsin IceCube Particle Astrophysics Center, University of Wisconsin, Madison, WI 53706, USA}
\author{Y.~Abdou}
\affiliation{Dept.~of Physics and Astronomy, University of Gent, B-9000 Gent, Belgium}
\author{M.~Ackermann}
\affiliation{DESY, D-15735 Zeuthen, Germany}
\author{J.~Adams}
\affiliation{Dept.~of Physics and Astronomy, University of Canterbury, Private Bag 4800, Christchurch, New Zealand}
\author{J.~A.~Aguilar}
\affiliation{D\'epartement de physique nucl\'eaire et corpusculaire, Universit\'e de Gen\`eve, CH-1211 Gen\`eve, Switzerland}
\author{M.~Ahlers}
\affiliation{Dept.~of Physics and Wisconsin IceCube Particle Astrophysics Center, University of Wisconsin, Madison, WI 53706, USA}
\author{D.~Altmann}
\affiliation{Institut f\"ur Physik, Humboldt-Universit\"at zu Berlin, D-12489 Berlin, Germany}
\author{J.~Auffenberg}
\affiliation{Dept.~of Physics and Wisconsin IceCube Particle Astrophysics Center, University of Wisconsin, Madison, WI 53706, USA}
\author{X.~Bai}
\thanks{Physics Department, South Dakota School of Mines and Technology, Rapid City, SD 57701, USA}
\affiliation{Bartol Research Institute and Department of Physics and Astronomy, University of Delaware, Newark, DE 19716, USA}
% ...

\date{\today}

\collaboration{IceCube Collaboration}
\noaffiliation

\begin{abstract}
We have performed a search for muon neutrinos from dark matter annihilation in the center of the Sun with the 79-string configuration of the IceCube neutrino telescope. For the first time, the DeepCore sub-array is included in the analysis, lowering the energy threshold and extending the search to the austral summer. The 317 days of data collected between June 2010 and May 2011 are consistent with the expected background from atmospheric muons and neutrinos. Upper limits are set on the dark matter annihilation rate, with conversions to limits on spin-dependent and spin-independent WIMP-proton cross-sections for WIMP masses in the range 20 - 5000\,GeV/c$^{2}$. These are the most stringent spin-dependent WIMP-proton cross-sections limits to date above 35\,GeV/c$^{2}$ for most WIMP models.
\end{abstract}

\maketitle
%

While the presence of dark matter (DM) in the universe has been inferred through its gravitational interactions, its nature remains a mystery.
One of the most promising and experimentally accessible candidates for DM are so-called Weakly Interacting Massive Particles (WIMPs)~\cite{WIMPRef}, predicted in extensions of the Standard Model of particle physics (SM).
DM may be captured in large celestial bodies like the Sun~\cite{GOULD} where self-annihilation to SM particles can result in a flux of high-energy neutrinos. These neutrinos can be searched for as a point-like source by IceCube~\cite{ic22,8year}. 
These indirect searches for DM are sensitive to the WIMP-proton scattering cross section, which initiates the capture process in the Sun. They complement direct DM searches on Earth as they scale with the averaged DM density along the solar circle and are more sensitive to low WIMP velocities~\cite{Bruch_darkDisc}. Indirect searches depend only weakly on the underlying WIMP velocity distribution~\cite{Carsten_Schrott} and we have chosen parameters to be conservative in our analysis.

In this work, we present new IceCube limits on dark matter captured by the Sun, with data taken in the 79-string configuration of the detector.
This analysis incorporates two significant additions compared to previous work. Firstly, we extend the search to the austral summer when the Sun is above the horizon. This doubles the livetime of the analysis, but imposes new challenges to reduce the downgoing atmospheric muon background. Secondly, we search for neutrinos from WIMPs with masses ($m_{\chi}$) as low as 20\,GeV/c$^{2}$ whereas past IceCube searches have only been sensitive above 50\,GeV/c$^{2}$.

The IceCube detector~\cite{icecube} is situated at the South Pole.
Digital Optical Modules (DOMs) arranged on vertical strings deep in the ice sheet record the Cherenkov light emitted by relativistic charged particles, including such created in neutrino interactions in the ice. The detection of photon yields and arrival times in DOMs allows the reconstruction of direction and energy of the secondaries.
This analysis used 317 live-days of data taken between June 2010 and May 2011. During this period, the detector was operating in its 79-string configuration, which includes six more densely instrumented strings in the center of the array, optimized for low energies. These strings feature reduced vertical spacing between DOMs and higher quantum efficiency photomultiplier tubes. Along with the seven surrounding regular strings, they form the DeepCore subarray~\cite{deepCore}.
Both the improvement in livetime and in energy threshold, which this analysis has achieved over previous IceCube analyses, can be attributed to the use of the DeepCore array.\\


The background in this search consists of muons and neutrinos created in cosmic ray interactions in the Earth's atmosphere. The dominant down-going muon component is simulated with \texttt{CORSIKA}~\cite{corsika}, including simulations of single and coincident air showers. The $\nu_{\mu}$ and $\nu_{e}$ components of the atmospheric spectrum are generated following the Honda flux model~\cite{honda2006}. For verification and cross-checks, a dedicated simulation of atmospheric $\nu$s below $200$\,GeV/c$^{2}$ is performed with \texttt{GENIE}~\cite{genie}. The background at final analysis level from solar atmospheric neutrinos, originating from cosmic ray interactions in the Sun's atmosphere, has been calculated to be of order 1 event, independent of the flux model~\cite{solarAtm1,solarAtm2,solarAtm3}. To reduce the dependence on simulation and associated systematic errors, we use off-source data to estimate the background at all analysis levels. Background simulation is merely used to verify accurate understanding of the detector. Off-source data consists of data recorded when the Sun was outside the respective analysis region.

Propagation of muons through the ice is simulated~\cite{mmc}, and transport of light from these particles to the DOMs is performed using direct photon tracking~\cite{ppc}, taking into account measured ice properties~\cite{ice}. Particle and photon propagation simulations at the lowest targeted energies below $50$\,GeV/c$^{2}$ have been independently verified using \texttt{GEANT4}~\cite{geant4}.

In this work the full dataset is split into three independent non-overlapping event selections; first into `summer' and `winter' seasons, when the Sun is above and below the horizon, respectively.
The `winter' dataset is further split into a low energy sample (WL), with focus on neutrino-induced muon tracks starting within DeepCore, and a higher energy sample (WH), aiming to select track-like events with no particular containment requirement. The `summer' selection is a dedicated low energy event sample (SL) for which the surrounding IceCube strings are used as an active muon veto in order to select neutrino-induced events starting within DeepCore. Separation into these samples is necessary owing to the different characteristics of the overwhelming down-going muon background within each dataset.
The event selection is carried out separately for each independent sample and the final search is conducted using a combined likelihood function.
In order to avoid potential bias, a strict blindness criterion is imposed by scrambling the azimuthal position of the Sun in data.
 
In IceCube, filters pre-select data to enhance the content of signal-like muon
events above the dominant atmospheric muon background. To increase the signal-to-background ratio, we only select events that pass any of three filters: the dedicated DeepCore low energy filter~\cite{deepCore} and two filters selecting muon-like events with an upwards pointing track reconstruction.  At this point, the dataset is split into the two seasonal streams, where September 22nd 2010 and  March 22nd 2011 mark the beginning and ending dates of the SL selection. 
We first discuss the additional `winter' cut selections:
Cuts are applied on the zenith angle and quality of the likelihood-based track reconstruction, on hit and string multiplicity, and on timing and topological variables.  For DeepCore contained events, the zenith acceptance region is extended to reflect the broadened signal point spread function at low energies.

The first data reduction is followed by additional processing, including an estimate of the angular uncertainty of the muon track fit.
Some signal neutrinos will arrive in coincidence with atmospheric backgrounds (~10\%). In order to retain these signal events, a set of topological criteria are applied to `split' these combined hit patterns into distinct sub-events.  These sub-events are then processed as above, and undergo all subsequent event selection in their own right.
Following the addition of events from splitting, the dataset is divided into independent low and high energy event samples.
For events to be included in the WL sample, we demand that the number of hit DOMs within DeepCore must be larger than outside.
Additionally, the number of outside hits must be less than seven. This ensures that events with a long lever arm and therefore good angular resolution are assigned to the complement sample. Events that fail the WL criteria are classified as WH events, and undergo a series of additional, stricter cuts on the same variables as in the initial event selection. WL events, conversely, undergo a veto cut, removing events with hits in the $10$ upper most layers of DOMs on the regular (non-DeepCore) strings.
\begin{figure}[t]
  \centering
  \includegraphics[width=0.48\textwidth]{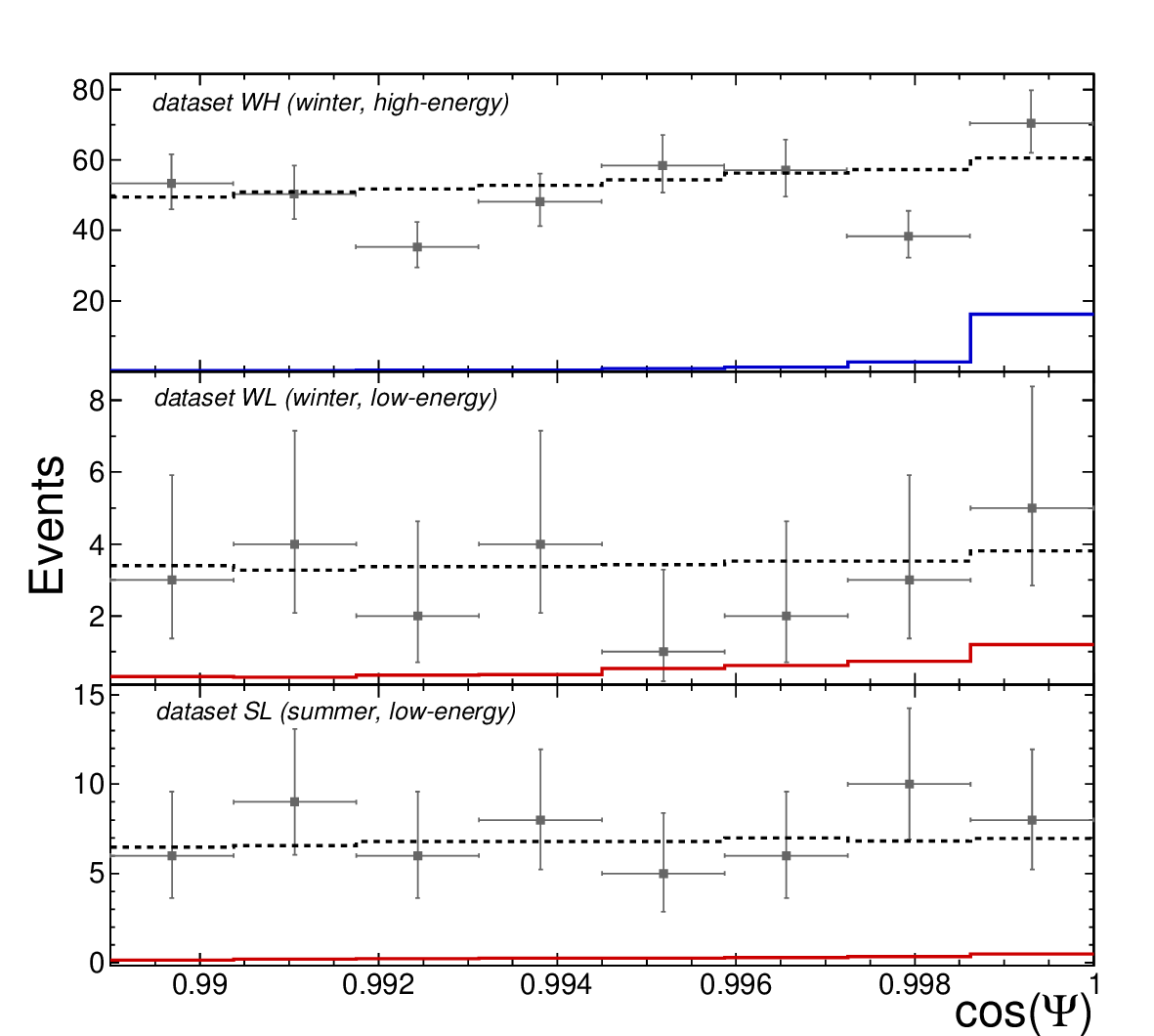}
  \caption{\label{fig:psi} Cosine of the angle between the reconstructed track and the direction of the Sun, $\Psi$, for observed events (squares) with one standard deviation error bars, and the atmospheric background expectation from atmospheric muons and neutrinos (dashed line). Also shown is a simulated signal (1\,TeV/c$^{2}$ hard for dataset WH, 50\,GeV/c$^{2}$ hard for datasets WL and SL) scaled to $\mu_{\mathrm{s}}^{90}$ (details in Table~\ref{tab:results}).}
\end{figure}
The final background reduction utilizes one Boosted Decision Tree (BDT)~\cite{TMVA} for each dataset to discriminate true up-going muon-like events from mis-reconstructed atmospheric muons. For training and testing an independent, high statistics, set of signal simulations is used and discarded afterwards. 
For background training this multivariate analysis uses one month of off-source data.
Through an iterative process, individual variables were removed and added and the performance of the BDT evaluated, until we arrived at a final set of 14 variables in the WH stream and 10 in the WL stream. All input distributions for simulated backgrounds and data are in good agreement. The selected variables describe both the quality of the track reconstruction and the time evolution of the pattern of hit DOMs and spatial positions within the detector.

The SL event sample uses a different set of cuts, because the dominant background is comprised of well-reconstructed down-going muons penetrating the detector. To reduce these backgrounds, we focus on low $m_{\chi}$ signals with a reconstructed neutrino interaction vertex inside the DeepCore fiducial volume. Selecting only these events, cuts are placed on the zenith angle of the track reconstruction, hit multiplicity, and vertical extension of the event. A 14 DOM layer top veto is imposed to reject down-going events.
Additionally, events are required to be DeepCore dominated (defined in the same way as for the `winter' analysis) and fulfill a tight hit-time containment criterion.
The final step in background rejection again consists of one BDT with 10 input variables.  These are selected using the same iterative selection process.
Track quality parameters yield less separation power within this down-going sample. As a result, the final BDT input observables mainly describe the degree of containment and the vertical and lateral extension of the event within the detector.  

The cut on the BDT score is optimized for each event selection to minimize the model rejection factor, MRF~\cite{MRF}, in the full likelihood analysis. Total signal cut-efficiencies range between 1-5\% for low $m_{\chi}$ signals and up to 30-40\% for high $m_{\chi}$.
The final step of the analysis is a likelihood ratio hypothesis test based on the values of the reconstructed angle to the Sun $\Psi$, using the Feldman-Cousins unified approach~\cite{fc}. This results in confidence intervals for the mean number of signal events, $\mu_{s}$.
The required probability densities for signal are computed from simulations, while for background they are based on real data events at the final selection level, with scrambled azimuth direction. 
A single result is calculated from all three data samples with a combined likelihood, constructed from the set of three independent probability distributions of signal and background, weighting each by the respective livetime and effective volume (see Ref.~\cite{8year} for details).

\begin{table*}
\caption{\label{tab:results} Results from the combination of the three independent datasets.  Upper 90\% limits on the number of signal events $\mu_{\mathrm{s}}^{90}$, the WIMP annihilation rate in the Sun $\Gamma_{\mathrm{A}}$, the muon flux $\Phi_{\mu}$ and neutrino flux $\Phi_{\nu}$, and the WIMP-proton scattering cross-sections (spin-independent, $\sigma_{SI,p}$, and spin-dependent, $\sigma_{SD,p}$), at the 90\% confidence level including systematic errors. The sensitivity $\overline{\Phi}_{\mu}$ (see text) is shown for comparison.}
\begin{ruledtabular}
 \begin{tabular}{cc|ccc|cccc}
 $m_{\chi}$&Channel&$\mu_{\mathrm{s}}^{90}$&$\Gamma_{\mathrm{A}}$&$\overline{\Phi}_{\mu}$&$\Phi_{\mu}$&$\Phi_{\nu}$&$\sigma_{SI,p}$&
$\sigma_{SD,p}$\\
 (GeV/c$^{2}$)&  &  & ($s^{-1}$) &($\mathrm{km}^{-2} \mathrm{y}^{-1}$)& ($\mathrm{km}^{-2} \mathrm{y}^{-1}$)& ($\mathrm{km}^{-2} \mathrm{y}^{-1}$)&($\mathrm{cm}^{2}$)&($\mathrm{cm}^{2}$)\\\hline
20 & $\tau^{+}\tau^{-}$ & $162$ & $2.46\times 10^{25}$ & $5.26\times 10^{4}$ & $9.27\times 10^{4}$ & $2.35\times 10^{15} $ & $1.08\times 10^{-40}$ & $1.29\times 10^{-38}$\\\hline
35 & $\tau^{+}\tau^{-}$ & $70.2$ & $1.03\times 10^{24}$ & $1.03\times 10^{4}$ & $1.21\times 10^{4}$ & $1.02 \times 10^{14}$ & $6.59\times 10^{-42}$ & $1.28\times 10^{-39}$\\
35 & $\mathrm{b}\bar{\mathrm{b}}$ & $128$ & $1.99\times 10^{26}$ & $5.63\times 10^{4}$ & $1.04\times 10^{5}$ & $6.29 \times 10^{15} $ & $1.28\times 10^{-39}$ & $2.49\times 10^{-37}$\\\hline
50 & $\tau^{+}\tau^{-}$ & $19.6$ & $1.20\times 10^{23}$ & $4.82\times 10^{3}$ & $2.84\times 10^{3}$ & $1.17 \times 10^{13} $ & $1.03\times 10^{-42}$ & $2.70\times 10^{-40}$\\
50 & $\mathrm{b}\bar{\mathrm{b}}$ & $55.2$ & $1.75\times 10^{25}$ & $2.06\times 10^{4}$ & $1.80\times 10^{4}$ & $5.64 \times 10^{14} $ & $1.51\times 10^{-40}$ & $3.96\times 10^{-38}$\\\hline
100 & $\mathrm{W}^{+}\mathrm{W}^{-}$ & $16.8$ & $3.35\times 10^{22}$ & $1.49\times 10^{3}$ & $1.19\times 10^{3}$ & $1.23 \times 10^{12} $ & $6.01\times 10^{-43}$ & $2.68\times 10^{-40}$\\
100 & $\mathrm{b}\bar{\mathrm{b}}$ & $28.9$ & $1.82\times 10^{24}$ & $7.57\times 10^{3}$ & $5.91\times 10^{3}$ & $6.34 \times 10^{13} $ & $3.30\times 10^{-41}$ & $1.47\times 10^{-38}$\\\hline
250 & $\mathrm{W}^{+}\mathrm{W}^{-}$ & $29.9$ & $2.85\times 10^{21}$ & $3.04\times 10^{2}$ & $4.15\times 10^{2}$ & $9.72 \times 10^{10} $ & $1.67\times 10^{-43}$ & $1.34\times 10^{-40}$\\
250 & $\mathrm{b}\bar{\mathrm{b}}$ & $19.8$ & $1.27\times 10^{23}$ & $1.85\times 10^{3}$ & $1.45\times 10^{3}$ & $4.59 \times 10^{12} $ & $7.37\times 10^{-42}$ & $5.90\times 10^{-39}$\\\hline
500 & $\mathrm{W}^{+}\mathrm{W}^{-}$ & $25.2$ & $8.57\times 10^{20}$ & $1.46\times 10^{2}$ & $2.23\times 10^{2}$ & $2.61 \times 10^{10} $ & $1.45\times 10^{-43}$ & $1.57\times 10^{-40}$\\
500 & $\mathrm{b}\bar{\mathrm{b}}$ & $30.6$ & $4.12\times 10^{22}$ & $8.53\times 10^{2}$ & $1.02\times 10^{3}$ & $1.52 \times 10^{12} $ & $6.98\times 10^{-42}$ & $7.56\times 10^{-39}$\\\hline
1000 & $\mathrm{W}^{+}\mathrm{W}^{-}$ & $23.4$ & $6.13\times 10^{20}$ & $1.19\times 10^{2}$ & $1.85\times 10^{2}$ & $1.62 \times 10^{10} $ & $3.46\times 10^{-43}$ & $4.48\times 10^{-40}$\\
1000 & $\mathrm{b}\bar{\mathrm{b}}$ & $30.4$ & $1.39\times 10^{22}$ & $4.33\times 10^{2}$ & $5.99\times 10^{2}$ & $5.23 \times 10^{11} $ & $7.75\times 10^{-42}$ & $1.00\times 10^{-38}$\\\hline
3000 & $\mathrm{W}^{+}\mathrm{W}^{-}$ & $22.2$ & $7.79\times 10^{20}$ & $1.09\times 10^{2}$ & $1.66\times 10^{2}$ & $1.65 \times 10^{10} $ & $3.44\times 10^{-42}$ & $5.02\times 10^{-39}$\\
3000 & $\mathrm{b}\bar{\mathrm{b}}$ & $26.1$ & $4.88\times 10^{21}$ & $2.52\times 10^{2}$ & $3.47\times 10^{2}$ & $1.89 \times 10^{11} $ & $2.17\times 10^{-41}$ & $3.16\times 10^{-38}$\\\hline
5000 & $\mathrm{W}^{+}\mathrm{W}^{-}$ & $22.8$ & $8.79\times 10^{20}$ & $1.01\times 10^{2}$ & $1.58\times 10^{2}$ & $1.77 \times 10^{10} $ & $1.06\times 10^{-41}$ & $1.59\times 10^{-38}$\\
5000 & $\mathrm{b}\bar{\mathrm{b}}$ & $26.4$ & $\footnote{Value has been corrected with respect to typo in published version of the paper.} 4.14\times 10^{21}$& $2.21\times 10^{2}$ & $3.26\times 10^{2}$ & $1.63 \times 10^{11} $ & $4.89\times 10^{-41}$ & $7.29\times 10^{-38}$\\
\end{tabular}
\end{ruledtabular}
\end{table*}

After unblinding the direction of the events in the final data samples, the observed distributions are compared to the expected background distributions from atmospheric muons and neutrinos, shown in Fig.~\ref{fig:psi}. The observed number of events from the direction of the Sun are consistent with the background-only hypothesis. Upper 90\% CL limits on $\mu_{\mathrm{s}}$ are calculated and listed for each signal hypothesis in Table~\ref{tab:results}.

The upper limit on $\mu_{\mathrm{s}}$ can be translated into a limit on the signal flux and annihilation rate in the Sun.
The effect of different sources of systematic uncertainties on signal flux expectations is calculated for three signal energy regions, defined in Table~\ref{tab:sys} by corresponding benchmark WIMP masses. Sources of uncertainties are divided into two classes; measurement and parameterization errors on cross sections and neutrino properties on the one hand and limitations in the detector simulation and uncertainties in detector calibrations on the other hand. The first class, Class-I, affects signal normalizations only, whereas the latter (Class-II) alters signal acceptance and introduces changes in the point spread function that is the basis for the likelihood analysis. Class-II uncertainties are evaluated using alternative signal simulations with varied calibration parameters, processed through the same analysis chain, and evaluated with the full multi-dataset combined likelihood. This procedure explicitly determines the systematic effect on $\mu_{s}$. 

Uncertainties in neutrino-nucleon cross-sections for signal simulations arise in the parameterization of the CTEQ6-DIS parton distribution functions as used in \texttt{nusigma}~\cite{nusigma}. In addition to this theoretical uncertainty on $\sigma_{\nu}$, the energy dependent error on the experimental $\sigma_{\nu}$-measurement~\cite{pdg} is included. The uncertainty in neutrino oscillation parameters used in signal flux calculations is investigated through variations of mixing parameters within the quoted $1 \sigma$ regions~\cite{pdg}. Here, the dominant effect results from the least constrained mixing angle, $\theta_{23}$, maximizing tau (dis)appearance within the expected flux expectation.

The second class of uncertainties includes absolute calibration and DOM to DOM variation of sensitivity, optical properties of the glacial ice, and photon propagation to the detector. The systematic uncertainties on absolute DOM sensitivity are evaluated with sets of signal simulations with an overall shift of 10\% in DOM efficiency. As baseline simulations do not account for varying relative DOM efficiency, dedicated signal simulations are performed with individual DOM efficiencies from a Gaussian fitted to the in-situ measured spread ($\sigma=0.087$) and centered around the nominal value. Optical properties of the glacial ice are measured~\cite{ice} and characterized in models that are parameterizations of the absorption and scattering coefficients as a function of depth and position in the detector. Two such models~\cite{ice,spice}, differing in parameterization techniques, are considered to bracket the uncertainty in light yield resulting from the ice description.
Individual uncertainties, listed in Table~\ref{tab:sys}, are added in quadrature to obtain the total systematic uncertainty for each benchmark mass region.\\

The upper limits on $\mu_{\mathrm{s}}$ for each signal hypothesis are then converted to limits on the neutrino to muon conversion rate and, through \texttt{DarkSUSY}~\cite{darksusy}, to limits on the WIMP annihilation rate in the Sun, $\Gamma_{\mathrm{A}}$. For better comparison to other experiments limits on the neutrino flux ($\Phi_{\nu}$) from the Sun, and the corresponding induced muon flux in the ice ($\Phi_{\mu}$), both integrated above 1\,GeV, are computed at the 90\% confidence level. These limits are listed in Table~\ref{tab:results}. Also specified is the median sensitivity, $\bar{\Phi}_{\mu}$, derived from simulations without signal.
\begin{figure}[!t]
  \centering
  \includegraphics[width=0.48\textwidth]{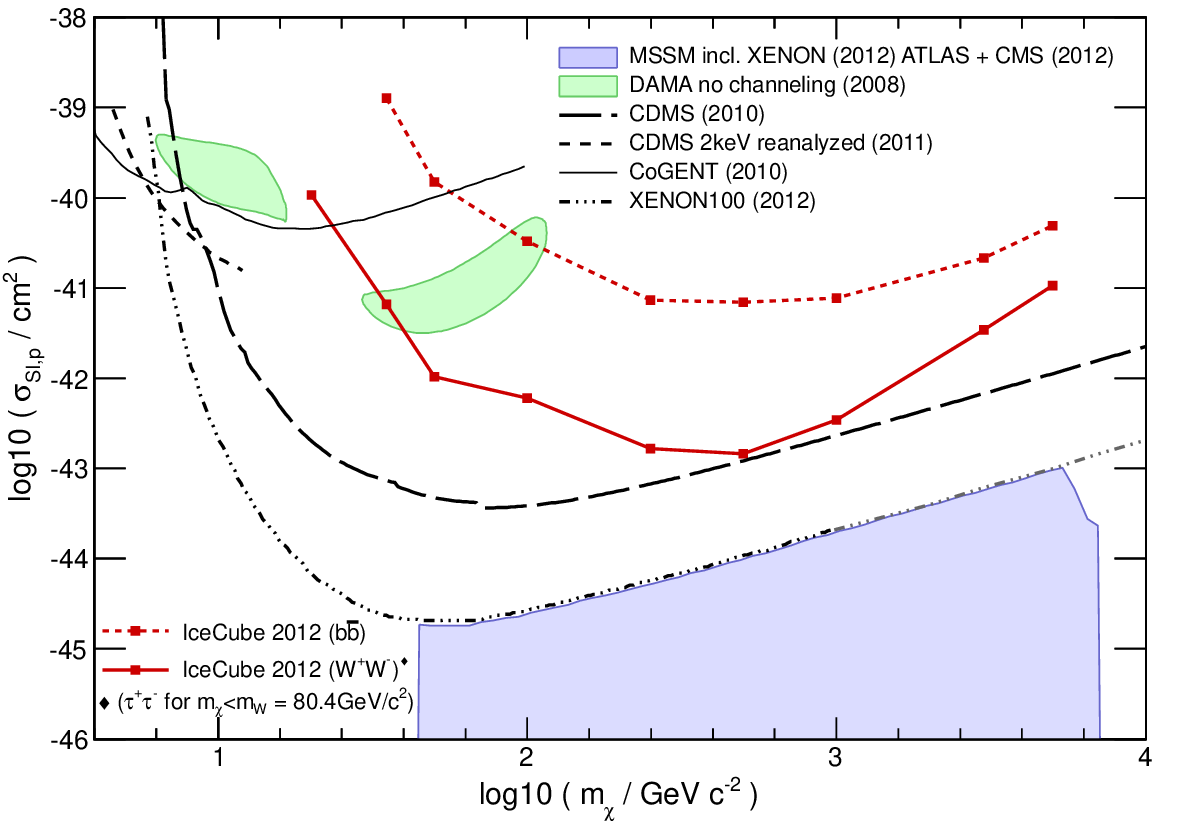}
  \includegraphics[width=0.48\textwidth]{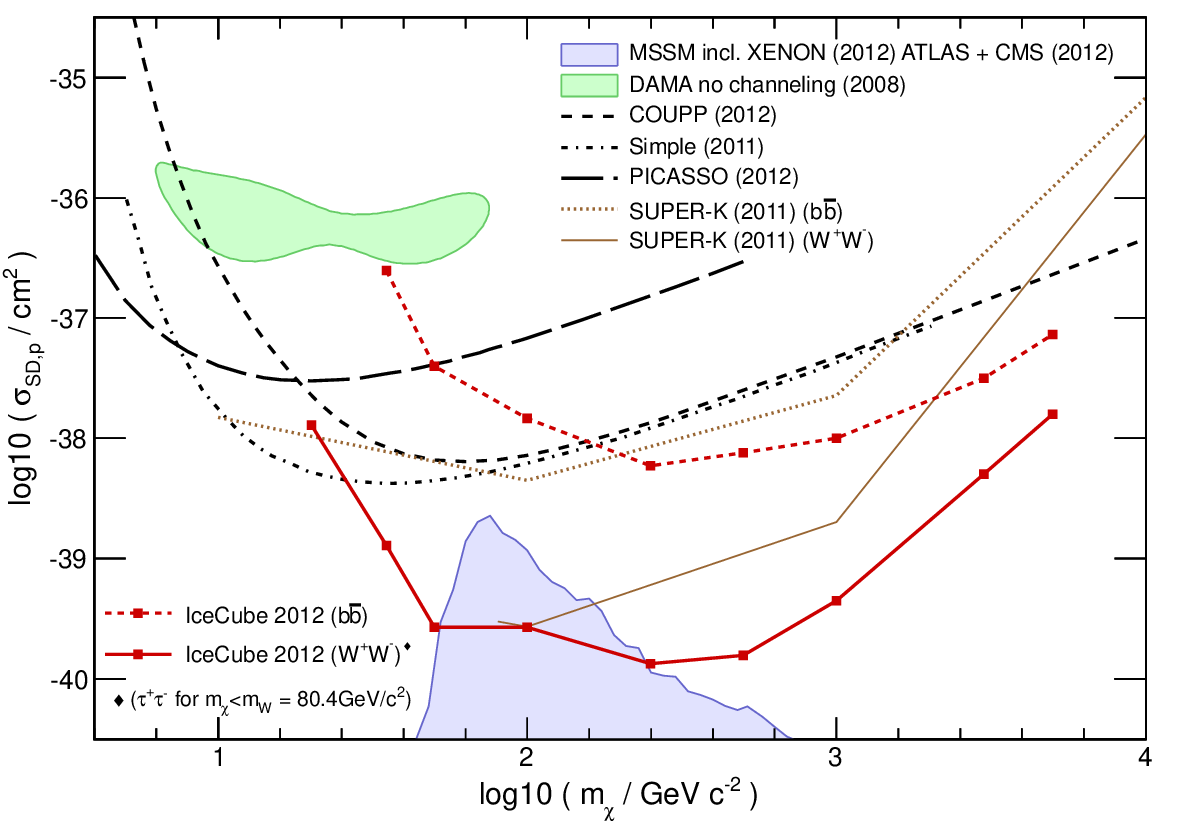}
  \caption{\label{fig:SI_SD}  90\% CL upper limits on $\sigma_{\it{SI,p}}$ (top figure) and $\sigma_{\it{SD,p}}$ (bottom figure) for hard and soft annihilation channels over a range of WIMP masses.  Systematic uncertainties are included.
  The shaded region represents an allowed MSSM parameter space (MSSM-25~\cite{MSSM25_paper}) taking into account recent accelerator~\cite{Razor}, cosmological and direct DM search constraints.
  Results from Super-K~\cite{superk}, COUPP(exponential model)~\cite{coupp} , PICASSO~\cite{picasso}, CDMS~\cite{cdms,cdmsLowE}, XENON100 (limits above 1\,TeV/c$^{2}$ from XENON100 Coll. private communication)~\cite{xenon}, CoGeNT~\cite{cogent}, Simple~\cite{simple} and DAMA~\cite{dama,dama2} are shown for comparison.}
\end{figure}
Under the assumption of equilibrium between WIMP capture and annihilation in the Sun, limits on $\Gamma_{\mathrm{A}}$ are converted into limits on the spin-dependent, $\sigma_{\it{SD,p}}$, and spin-independent, $\sigma_{\it{SI,p}}$, WIMP-proton scattering cross-sections, using the method from Ref~\cite{JCAPconversion}.
Results are listed in Table~\ref{tab:results} and shown in Fig.~\ref{fig:SI_SD} together with other experimental limits~\cite{superk,coupp,picasso,cdms,cdmsLowE,xenon,cogent,simple,dama,dama2}.  We assume a standard DM halo with a local density of $0.3$\,GeV/cm$^{3}$~\cite{pdg} and a Maxwellian WIMP velocity distribution with an RMS velocity of $270$\,km/s. We do not include the detailed effects of diffusion and planets upon the capture rate, as the simple free-space approximation~\cite{GOULD} included in \texttt{DarkSUSY} is found to be accurate~\cite{JoakimSophiaSuncapture}. Limits on the WIMP-nucleon scattering cross section can also be deduced from limits on mono-jet and mono-photon signals at hadron colliders, but these depend strongly on the choice of the underlying effective theory and mediator masses~\cite{LHC_mono_searches,LHC_mono_searches2,LHC_mono_searches3}, and consequently not included in Fig.~\ref{fig:SI_SD}.
\begin{table}
  \caption{\label{tab:sys} Systematic errors on signal flux expectations in percent. Class-II uncertainties marked $^{*}$}
  \centering
  \begin{tabular}{c|c|c|c}
    \hline\hline
    Source & \multicolumn{3}{c}{mass ranges (GeV/c$^{2}$)}\\
    & \,$<35$\, & 35\,-100 & \,$>100$\, \\\hline
    $\nu$ oscillations & 6 & 6  & 6  \\
    $\nu$-nucleon cross-section & 7 & 5.5  &  3.5 \\
    $\mu$-propagation in ice & $<$1 &  $<$1 &  $<$1 \\
    Time, position calibration & 5 & 5 & 5 \\
    DOM sensitivity spread$^{*}$ & 6 & 3 &  10 \\
    Photon propagation in ice$^{*}$ & 15 & 10  &  5 \\
    Absolute DOM efficiency$^{*}$ & 50 & 20  &  15 \\
    \hline
    Total uncertainty & 54 & 25 & 21 \\
    \hline\hline
  \end{tabular}
\end{table}

In conclusion, we have presented the most stringent limits to date on the spin-dependent WIMP-proton cross-section for WIMPs annihilating into $W^{+}W^{-}$ or $\tau^{+}\tau^{-}$ with masses above 35\,GeV/c$^{2}$.
With this dataset, we have demonstrated for the first time the ability of IceCube to probe WIMP masses below 50\,GeV/c$^{2}$. This has been accomplished through effective use of the DeepCore sub-array. Furthermore, we have accessed the southern sky for the first time by incorporating strong vetos against the large atmospheric muon backgrounds. The added livetime has been shown to improve the presented limits.
IceCube has now achieved limits that strongly constrain dark matter models and that will impact global fits of the allowed dark matter parameter space. This impact will only increase in the future, as analysis techniques improve and detector livetime increases. 

\begin{acknowledgments}
We thank H.~Silverwood for his support on SUSY model scans.
We acknowledge the support from the following agencies:
U.S. National Science Foundation-Office of Polar Programs,
U.S. National Science Foundation-Physics Division,
University of Wisconsin Alumni Research Foundation,
the Grid Laboratory Of Wisconsin (GLOW) grid infrastructure at the University of Wisconsin - Madison, the Open Science Grid (OSG) grid infrastructure;
U.S. Department of Energy, and National Energy Research Scientific Computing Center,
the Louisiana Optical Network Initiative (LONI) grid computing resources;
National Science and Engineering Research Council of Canada;
Swedish Research Council,
Swedish Polar Research Secretariat,
Swedish National Infrastructure for Computing (SNIC),
and Knut and Alice Wallenberg Foundation, Sweden;
German Ministry for Education and Research (BMBF),
Deutsche Forschungsgemeinschaft (DFG),
Helmholtz Alliance for Astroparticle Physics (HAP),
Research Department of Plasmas with Complex Interactions (Bochum), Germany;
Fund for Scientific Research (FNRS-FWO),
FWO Odysseus programme,
Flanders Institute to encourage scientific and technological research in industry (IWT),
Belgian Federal Science Policy Office (Belspo);
University of Oxford, United Kingdom;
Marsden Fund, New Zealand;
Australian Research Council;
Japan Society for Promotion of Science (JSPS);
the Swiss National Science Foundation (SNSF), Switzerland.
\end{acknowledgments}


\begin{thebibliography}{99}
  \bibitem{WIMPRef}
G. Bertone, D. Hooper and J. Silk, Phys. Rept. {\bf405}, 279 (2005).

  \bibitem{GOULD}
A.~Gould, Ap.~J. {\bf328}, 919 (1988);
J.~Silk \etal Phys. Rev. Lett. {\bf55}, 257 (1985);
T.~K.~Gaisser \etal Phys. Rev. {\bf D34}, 2206 (1986).

  \bibitem{ic22}
R.~Abbasi \etal Phys. Rev. Lett. {\bf 102}, 201302 (2009).

  \bibitem{8year}
R.~Abbasi \etal Phys. Rev. {\bf D85}, 042002 (2012).

  \bibitem{Bruch_darkDisc}
T.~Bruch \etal Phys. Letters {\bf B674}, 250 (2009).

  \bibitem{Carsten_Schrott}
C.~Rott \etal JCAP {\bf 09}, 029 (2011).

  \bibitem{icecube}
R.~Abbasi \etal Nucl. Instrum. Methods {\bf A618}, 139 (2010);
R.~Abbasi \etal Nucl. Instrum. Methods {\bf A601}, 294 (2009).

  \bibitem{deepCore}
R.~Abbasi \etal Astropart. Phys. {\bf 35}, 615 (2012).

  \bibitem{darksusy}
P. Gondolo \etal JCAP {\bf 07}, 008  (2004).

  \bibitem{wimpsim}
M.~Blennow \etal JCAP {\bf01}, 021 (2008).

  \bibitem{corsika}
D. Heck \etal FZKA report 6019, (1998). \\
{(\url{http://www-ik.fzk.de/corsika})}.

  \bibitem{honda2006}
T.~K.~Gaisser and M.~Honda, Rev. of Nucl. {\&} Part. S. {\bf 52}, 153 (2002);
M.~Honda \etal  Phys. Rev. {\bf D75}, 043006 (2007).

  \bibitem{genie}
C.~Andreopoulos \etal Nucl. Instrum. Methods {\bf A614} 87 (2010). 

  \bibitem{solarAtm1}
G.~Ingelman and M.~Thunman, Phys. Rev. {\bf D54}, 4385 (1996).

  \bibitem{solarAtm2}
G.~L.~Fogli \etal Phys. Rev. {\bf D74}, 3004 (2006).

  \bibitem{solarAtm3}
C.~Hettlage, K.~Mannheim and J.~G.~Learned, Astropart. Phys. {\bf 13}, 45 (2000).

  \bibitem{geant4}
S.~Agostinelli \etal Nucl. Instrum. Methods {\bf A506}, 250  (2003).

  \bibitem{mmc}
D. Chirkin and W. Rhode, Proc. of the 27th Intl. Cosmic Ray Conference, Hamburg, Germany, 2001, HE2.02, 1017, and hep-ph/0407075.

  \bibitem{ppc}
D.~Chirkin, Nucl. Instrum. Methods {\bf A} (2012);
{(\url{http://dx.doi.org/10.1016/j.nima.2012.11.170})}

  \bibitem{ice}
M.~Ackermann \etal  J. Geophys. Res., {\bf 111}, D13203  (2006).

\bibitem{spice}
The IceCube Collaboration, proceedings ICRC 2011, arXiv:1111.2731 (2011).

  \bibitem{nusigma}
J.~Edsj\"o, The nusigma neutrino-nucleon scattering Monte Carlo (2007). {(\url{http://www.physto.se/~edsjo})} 

  \bibitem{pdg}
Particle Data Group, PR {\bf D86}, 010001 (2012).

  \bibitem{fc}
G.~J.~Feldman and R. D. Cousins, Phys. Rev. {\bf D57}, 3873  (1998). 

  \bibitem{MRF}
G.~Hill and K.~Rawlins, Astropart. Phys. {\bf 19} 393 (2003).

  \bibitem{MSSM25_paper}
H.~Silverwood \etal arXiv:1210.0844v1 (2012).

  \bibitem{Razor}
The ATLAS Collaboration, ATLAS-CONF-2012-033 (2012);
The CMS Collaboration, CMS-PAS-SUS-11-024 (2012);
The CMS Collaboration, arXiv:1207.1798 (2012).

  \bibitem{superk}
T.~Tanaka \etal  Astrophys. J. {\bf 742}, 78 (2011). 

  \bibitem{coupp}
E.~Behnke \etal Phys. Rev. {\bf D86}, 052001 (2012).


  \bibitem{picasso}
S.~Archambault \etal Phys. Lett. {\bf B711}, 153 (2012).

  \bibitem{cdms}
Z.~Ahmed \etal Science {\bf327}, 5973:1619-1621, (2011).

  \bibitem{cdmsLowE}
Z.~Ahmed \etal Phys. Rev. Lett. {\bf 106} 131302 (2011).

  \bibitem{dama}
R.~Bernabei \etal European Phys. J. {\bf C56}, 3 (2008).

  \bibitem{dama2}
C.~Savage \etal JCAP {\bf 04}, 010 (2009).

  \bibitem{cogent}
C.~E.~Aalseth \etal Phys. Rev. Lett. {\bf 106}, 131301 (2011). 

  \bibitem{xenon}
E.~Aprile \etal Phys. Rev. Lett. {\bf 109}, 181301 (2012);
M.~Schumann, analysis coordinator, for the XENON100 Coll., private communication.

  \bibitem{simple}
M.~Felizardo \etal Phys. Rev. Lett. {\bf108}, 201302 (2012).

  \bibitem{TMVA}
H.~Voss \etal PoS ACAT {\bf040} (2007),\\ and arXiv:physics/0703039v5 (2009).

  \bibitem{JCAPconversion}
G.~Wikstr\"om and J.~Edsj\"o, JCAP {\bf0904}, 009 (2009).

  \bibitem{JoakimSophiaSuncapture}
S.~Sivertsson and J.~Edsj\"o, Phys. Rev. {\bf D85}, 123514 (2012).

  \bibitem{LHC_mono_searches}
Y. Bai, P. J. Fox, and R. Harnik, JHEP {\bf 1012}, 048 (2010).

  \bibitem{LHC_mono_searches2}
J.~Goodman \etal Phys. Lett. B {\bf695}, 185 (2011).

  \bibitem{LHC_mono_searches3}
P.~J.~Fox \etal Phys. Rev. D {\bf85}, 056011 (2012).
\end{thebibliography}
\end{document}